\documentclass[final]{siamltex}

\renewcommand{\baselinestretch}{1.08}
\usepackage{cite}
\usepackage{amsmath,amssymb,amsfonts}
\usepackage{algorithmic}
\usepackage{graphicx}
\usepackage{textcomp}
\usepackage{xcolor}

\usepackage{multirow}

\title{Hierarchical Block Multi-Color Ordering: A New Parallel Ordering Method for Vectorization and Parallelization of the Sparse Triangular Solver in the ICCG Method}

\author{Takeshi Iwashita\footnotemark[2] , Senxi Li\footnotemark[3] \and Takeshi Fukaya\footnotemark[3]}

\newcommand{\0}{{\bf 0}}

\newcommand{\bbb}{\mbox{\boldmath $b$}}

\newcommand{\qqq}{\mbox{\boldmath $q$}}
\newcommand{\rrr}{\mbox{\boldmath $r$}}

\newcommand{\ttt}{\mbox{\boldmath $t$}}

\newcommand{\xxx}{\mbox{\boldmath $x$}}
\newcommand{\yyy}{\mbox{\boldmath $y$}}
\newcommand{\zzz}{\mbox{\boldmath $z$}}
\newcommand{\AAA}{\mbox{\boldmath $A$}}
\newcommand{\BBB}{\mbox{\boldmath $B$}}
\newcommand{\CCC}{\mbox{\boldmath $C$}}
\newcommand{\DDD}{\mbox{\boldmath $D$}}
\newcommand{\EEE}{\mbox{\boldmath $E$}}

\newcommand{\JJJ}{\mbox{\boldmath $J$}}

\newcommand{\LLL}{\mbox{\boldmath $L$}}

\newcommand{\PPP}{\mbox{\boldmath $P$}}

\newcommand{\UUU}{\mbox{\boldmath $U$}}

\newcommand{\beq}{\begin{equation}}
\newcommand{\eeq}[1]{\label{#1} \end{equation}}
\newcommand{\beqa}{\begin{eqnarray}}
\newcommand{\eeqa}[1]{\label{#1} \end{eqnarray}}
\newcommand{\bmat}[1]{\left ( \begin{array}{#1}}
\newcommand{\emat}{\end{array} \right )}

\begin{document}

\maketitle

\renewcommand{\thefootnote}{\fnsymbol{footnote}}

\footnotetext[2]{Information Initiative Center, Hokkaido University, Sapporo, Japan. (iwashita@iic.hokudai.ac.jp, fukaya@iic.hokudai.ac.jp)}
\footnotetext[3]{Graduate School of Information Science and Technology, The University of Tokyo, Tokyo, Japan. (lisenxi137@gmail.com)}

%
\begin{abstract}
In this paper, we propose a new parallel ordering method to vectorize and parallelize the sparse triangular solver, which is called hierarchical block multi-color ordering. 
In this method, the parallel forward and backward substitutions can be vectorized while preserving the advantages of block multi-color ordering, that is, fast convergence and fewer thread synchronizations. 
To evaluate the proposed method in a parallel ICCG (Incomplete Cholesky Conjugate Gradient) solver, numerical tests were conducted using five test matrices on three types of computational nodes.
The numerical results indicate that the proposed method outperforms the conventional block and nodal multi-color ordering methods in 13 out of 15 test cases, which confirms the effectiveness of the method.
\end{abstract}

%
%
%
%
%

\begin {keywords}
Sparse triangular solver, Parallel ordering, ICCG method, SIMD vectorization, Multithreading
\end{keywords}

%

%

\section{Introduction}
A sparse triangular solver is an important computational kernel for an iterative linear solver in various numerical simulations.
It is the main component of the Gauss--Seidel (GS)  smoother,  SOR method and IC/ILU preconditioning, which are used as building blocks in various computational science or engineering analyses~\cite{Iterativereview, saad, meurant}.
Therefore, the development of a fast multithreaded sparse triangular solver is essential to accelerate these analyses when conducted on not only a single computing node but also a large-scale cluster system of nodes. 
For example, the performance of the solver significantly influences the total simulation time of large-scale partial differential equation  analysis using a multigrid solver with the GS, IC, or ILU smoother~\cite{Wallin, Buckeridge}.
However, it is well known that the sparse triangular solver, which consists of forward and backward substitutions, cannot be straightforwardly parallelized~\cite{Dongarra, Dongarra2}. 
Thus, in this paper, we discuss an effective approach to developing a high-performance multithreaded sparse triangular solver.
%
%
%
%
%
%
%
%
%
%
%

There are various methods for parallelizing a sparse triangular solver or its related techniques, and we focus on the parallel ordering (reordering) method, which is one of the most common methods for parallelization of a sparse triangular solver.
There are several well-known orderings, such as dissection and domain decomposition orderings, but multi-color ordering is the most commonly used technique. 
It has been used in various applications to parallelize, for example, the ICCG method.
However, it is well known that the multi-color ordering entails a trade-off problem between convergence and the number of synchronizations~\cite{doi4}. 
An increase in the number of colors typically results in better convergence, but it also leads to an increase in the number of synchronizations, which is proportional to the number of colors. 
The trade-off problem between convergence and parallelism is a common issue for parallel ordering techniques~\cite{Duff}.




One of the solutions for the above trade-off problem is {\it block} multi-coloring. 
In this method, multi-color ordering is applied to blocks of unknowns.
The technique has been investigated in several contexts.
The concept of block coloring or block independent sets can be seen in \cite{saad}.
In an early work~\cite{SOR-bc}, it is discussed for the parallel SOR method.
For parallelization of the IC/ILU preconditioned iterative solver, it was first investigated in a finite difference method, that is, structured grid analysis~\cite{brb1, siam-iwa}. In this research, block coloring proved to be effective for improving convergence without increasing thread synchronization.
Following on from these research activities, the algebraic block multi-color ordering method was introduced for a general sparse linear system in \cite{IPDPS2012}.
Although there are various options for coloring or blocking methods~\cite{multi,amc}, this technique has been used in various applications because of its advantages in terms of convergence, data locality, and the number of synchronizations~\cite{semba,tsuburaya}. Particularly, several high-performance implementations of the HPCG benchmark adopt the technique, which shows the effectiveness of the method in a fast multigrid solver with the parallel GS smoother~\cite{intel,HPCG-kumahata,PEZY,HPCG-ICPP,arm}. 
However, the block multi-coloring method has a drawback in its implementation using SIMD vectorization.
The calculations in the innermost loop for the parallel substitutions are performed sequentially, which prevents the efficient use of SIMD instructions.

Because the sparse triangular solver is a memory-intensive kernel, its performance on previous computer was not substantially affected by the use of SIMD instructions.
However, to increase the floating-point performance,
recent processors enhance the SIMD instructions and their SIMD width (vector length) is becoming large.
For example, Intel Xeon (Skylake)~\cite{Skylake}, Intel Xeon Phi~\cite{KNL}, and Fujitsu A64FX (ARM SVE) ~\cite{post-K} processors are equipped with 512-bit SIMD instructions. We note that ARM SVE supports at most a 2,048 vector length~\cite{armSVE}.
Considering this trend of processors, we aim to develop a parallel sparse triangular solver in which both multithreading and SIMD vectorization are efficiently used.

In this paper, we propose a new parallel ordering technique in which SIMD vectorization can be used and the advantages of block multi-color ordering, that is, fast convergence and fewer synchronizations, are preserved.
The technique is called ``hierarchical block multi-color ordering'' and it has a mathematically equivalent solution process (convergence) to block multi-color ordering. Moreover, the number of synchronizations in the multithreaded substitutions is the same as that of block multi-color ordering. We conduct five numerical tests using finite element electromagnetic field analysis code and matrix data obtained from a matrix collection, and confirm the effectiveness of the proposed method in the context of the parallel ICCG solver.

\section{Sparse Triangular Solver}
In this paper, we consider the following $n$-dimensional linear system of equations:
\beq
\AAA \xxx = \bbb.
\eeq{org}
We discuss the case in which the linear system (\ref{org}) is solved using an iterative linear solver involving IC(0)/ILU(0) preconditioning, the Gauss-Seidel (GS) smoother, or the SOR method.
When we discuss a parallel ICCG (precisely IC(0)-CG) solver for (\ref{org}), we assume that coefficient matrix $\AAA$ is symmetric and positive or semi-positive definite. 
For the parallelization of the iterative solver that we consider, the most problematic part is in the sparse triangular solver kernel.
For example, in an IC/ILU preconditioned Krylov subspace iterative solver, the other computational kernels consist of an inner product, matrix-vector multiplication, and vector updates, which can be parallelized straightforwardly.
%
%
The sparse triangular solver kernel is given by following forward and backward substitutions:
\beq
\yyy = \LLL^{-1} \rrr,
\eeq{for}
and
\beq
\zzz = \UUU^{-1} \yyy,
\eeq{back}
where $\rrr$, $\yyy$, and $\zzz$ are $n$-dimensional vectors. Matrices $\LLL$ and $\UUU$ are, respectively, lower and upper triangular matrices with the same nonzero patterns as the lower and upper triangular parts of $\AAA$. In ILU (IC) preconditioning, the preconditioning step is given by (\ref{for}) and (\ref{back}), and triangular matrices $\LLL$ and $\UUU$ are derived from the following incomplete factorization:
\beq
\AAA \simeq \LLL \UUU.
\eeq{ILU}
The iteration steps in the GS and SOR methods (smoothers) can be expressed by similar substitutions. 
The substitution is an inherently sequential process, and it cannot be parallelized (multithreaded) straightforwardly.


\section{Parallel Ordering Method}
A parallel ordering (reordering) method is one of the most popular parallelization methods for a sparse triangular solver, that is, forward and backward substitutions.
It transforms the coefficient matrix into an appropriate form for parallel processing by reordering the unknowns or their indices. 
Let the reordered linear system of (\ref{org}) be denoted by
\beq
\bar{\AAA} \bar{\xxx} = \bar{\bbb}.
\eeq{reordered} 
Then, the {\it reordering} is given by the transformation:
\beq
\bar{\xxx} = \PPP_{\pi} \xxx,
\eeq{xbar}
where $\PPP_{\pi}$ is a permutation matrix.
When we consider index set $I=\{1, 2, \ldots, n\}$ that corresponds to the index of each unknown, 
the reordering is the permutation of the elements of $I$.
In the present paper, the reordering function of the index is denoted by $\pi$; that is, the $i$-th unknown of the original system is moved to the $\pi(i)$-th unknown of the reordered system.
In the reordering technique, the coefficient matrix and right-hand side are given as follows:
\beq
\bar{\AAA}=\PPP_{\pi} \AAA \PPP_{\pi}^{\top}, \quad \bar{\bbb}=\PPP_{\pi} \bbb.
\eeq{AB}

\subsection{Equivalence of orderings}
We consider the case in which two linear systems, (\ref{org}) and (\ref{reordered}), are solved using an identical iterative method.
The approximate solution vector at the $j$-th iteration for (\ref{org}) and that for (\ref{reordered}) are denoted by $\xxx^{(j)}$, and $\bar{\xxx}^{(j)}$, respectively.
If it holds that 
\beq
\bar{\xxx}^{(j)} = \PPP_{\pi} \xxx^{(j)}
\eeq{xj}  
at every $j$-th step under the setting $\bar{\xxx}^{(0)} = \PPP_{\pi} \xxx^{(0)}$ for initial guesses, then we can say that these two solution processes are $equivalent$. For example, in the Jacobi method and most Krylov subspace methods, reordering does not affect convergence; that is, the solution process for any reordered system is (mathematically) equivalent to that for the original system.
However, in the case of the iterative solver that we consider in this paper, such as the IC/ILU preconditioned iterative solver, the solution processes are typically inequivalent. 
This is because of the sequentiality involved in the triangular solver (substitutions).
However, there are special cases in which the reordered system has an equivalent solution process to the original system. 
In these cases, we say that two (original and new) orderings are equivalent or $\pi$ is an equivalent reordering.

We define the equivalence of two orderings as follows:
In the GS and SOR methods, equivalence is given by (\ref{xj}) under the proper setting of the initial guess. 
In IC(0)/ILU(0) preconditioning, equivalence is given as follows:
We denote the incomplete factorization matrices of $\bar{\AAA}$ by $\bar{\LLL}$ and $\bar{\UUU}$. Moreover, the preconditioning step of the reordered linear system is given by $\bar{\zzz} = (\bar{\LLL} \bar{\UUU})^{-1} \bar{\rrr}$.
If $\bar{\zzz}=\PPP_{\pi} \zzz$ is satisfied under $\bar{\rrr} = \PPP_{\pi} \rrr$, then we say that the orderings are equivalent.
For example, the ICCG (IC(0)-CG) method exhibits an equivalent solution process for both original and reordered linear systems when the orderings are equivalent.

\begin{figure}[tbp]
\centering
\includegraphics[scale=0.6,clip, bb= 10 150 620 480]{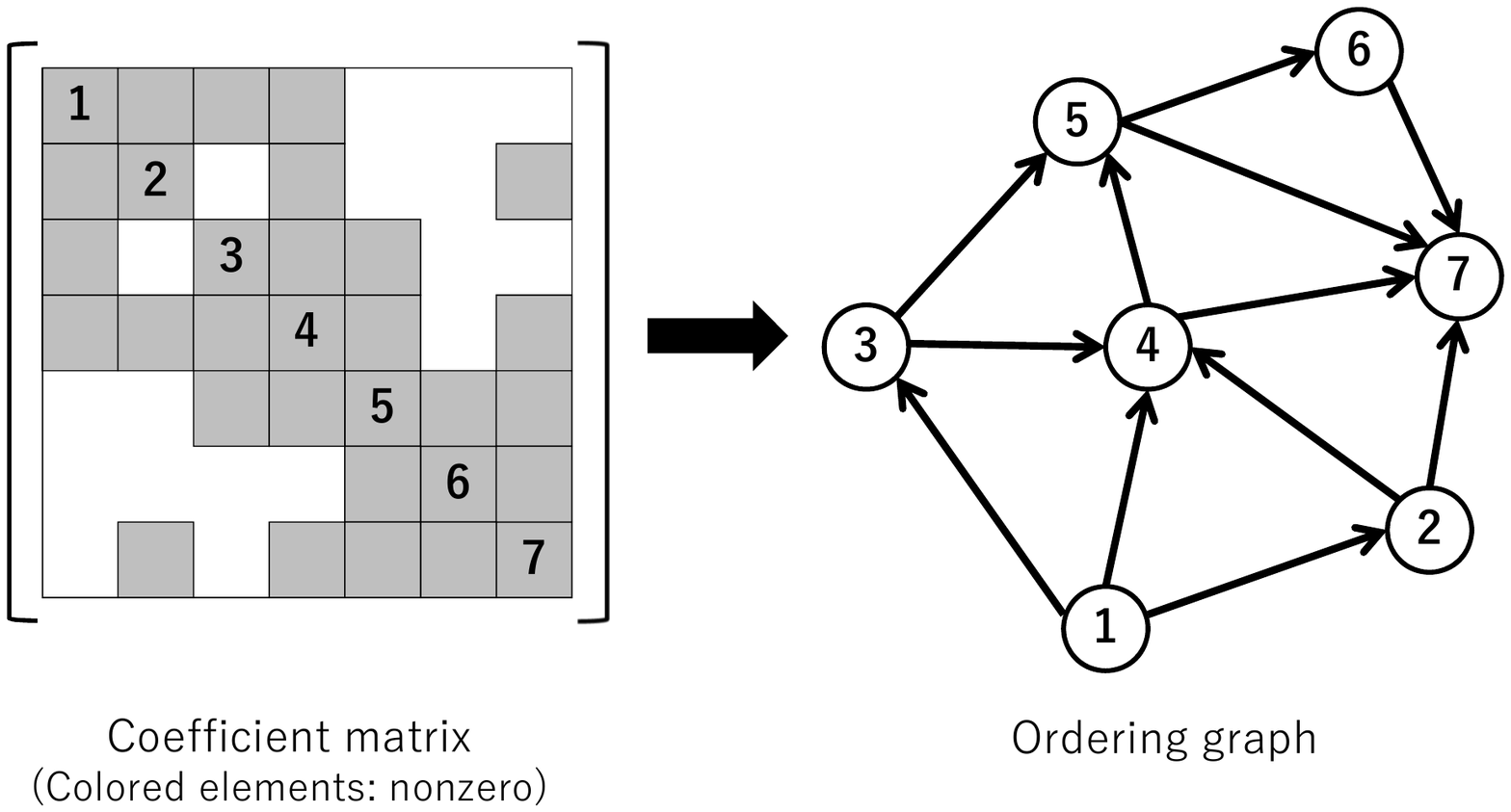}
\caption{Example of an ordering graph} 
\label{og}
\end{figure}

The condition for equivalent reordering is given as follows:
When the following ER condition is satisfied, $\pi$ is the equivalent reordering.
\begin{quote}
{\it ER (Equivalent Reordering) Condition} --- 
\begin{eqnarray}
\forall i_{1}, i_{2} \in I \ {\rm such} \ {\rm that} \  a_{i_{1}, i_{2}} \neq 0 \ \vee \ a_{i_{2}, i_{1}} \neq 0, \nonumber \\ 
\mbox{sgn} (i_{1}-i_{2}) = \mbox{sgn} (\pi (i_{1})-\pi(i_{2})),
\label{equivalence} 
\end{eqnarray}
\end{quote}
where $a_{i_{1}, i_{2}}$ denotes the $i_{1}$-th row $i_{2}$-th column element of $\AAA$.
%
%
%
%
%
For a further explanation, we introduce an ordering graph, which is the directed graph that corresponds to the coefficient matrix. 
Each node of the graph corresponds to an unknown or its index. An edge between two nodes $i_{1}$ and $i_{2}$ exists only when the $i_{1}$-th row $i_{2}$-th column element or $i_{2}$-th row $i_{1}$-th column element is nonzero.
The direction of the edge (arrow) shows the order of two nodes.
Figure \ref{og} shows an example of the ordering graph.
Using the ordering graph, (\ref{equivalence}) can be rewritten as the statement that the new and original orderings have the same ordering graph.
In \cite{doi3}, the authors stated that the ordering graph provides a unique class of mutually equivalent orderings. 
In the appendix, we provide a proof sketch of the relationship between (\ref{equivalence}) and the equivalence of orderings.
%
%


%

\section{Hierarchical Block Multi-Color Ordering Method }
In this paper, we propose a new parallel ordering method for the vectorization and parallelization of  a sparse triangular solver. 
Additionally, the proposed ordering is intended to inherit the advantages of convergence, number of synchronizations, and data locality from block multi-color ordering (BMC).
The proposed parallel ordering is called hierarchical block multi-color ordering (HBMC), which is equivalent to BMC in terms of convergence.

In the technique, we first order the unknowns by using  BMC, and then reorder them again. 
We focus on the explanation of the secondary reordering because we use the conventional algorithm shown in \cite{IPDPS2012} for the application of BMC.
Therefore, the original linear system based on BMC is written as (\ref{org}) and secondary reordering is denoted by $\pi$. Thus, the final reordered linear system based on HBMC is given by (\ref{reordered}).

\subsection{Block multi-color ordering (BMC)}
\begin{figure}[tbp]
\centering
\includegraphics[scale=0.65,clip, bb= 105 165 550 480]{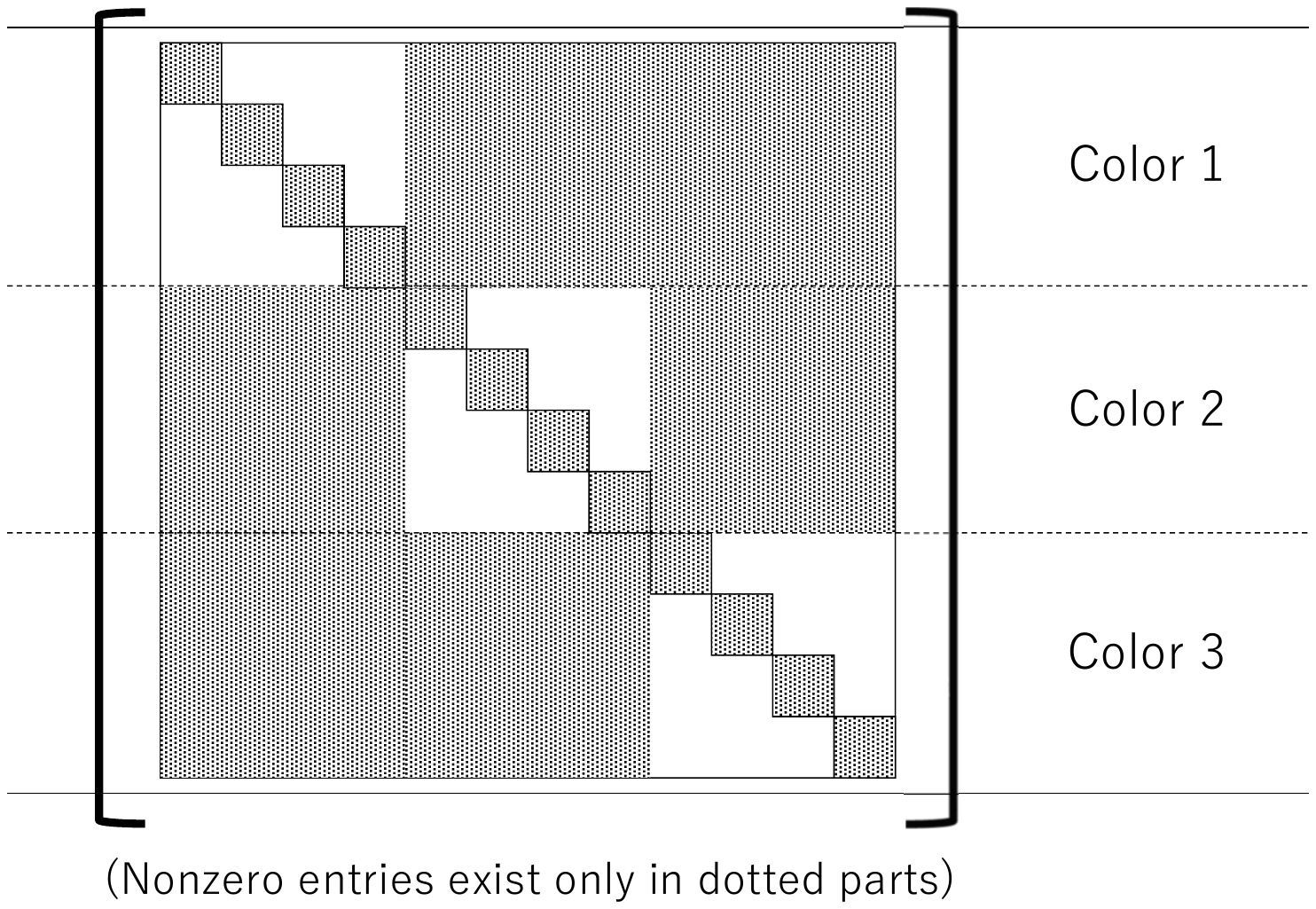}
\caption{Coefficient matrix based on BMC} 
\label{BMC-1}
\end{figure}

In this subsection, we briefly introduce BMC and some notation required for the explanation of HBMC.
In BMC, all unknowns are divided into blocks of the same size, and the multi-color ordering is applied to the blocks.
Because blocks that have an identical color are mutually independent, the forward and backward substitutions are parallelized based on the blocks in each color. The number of (thread) synchronizations of parallelized (multithreaded) substitution is given by $n_{c}-1$, where $n_{c}$ is the number of colors.
Figure \ref{BMC-1} shows the coefficient matrix that results from BMC.

\begin{figure}[tbp]
\centering
\includegraphics[scale=0.65,clip, bb= 33 76 615 420]{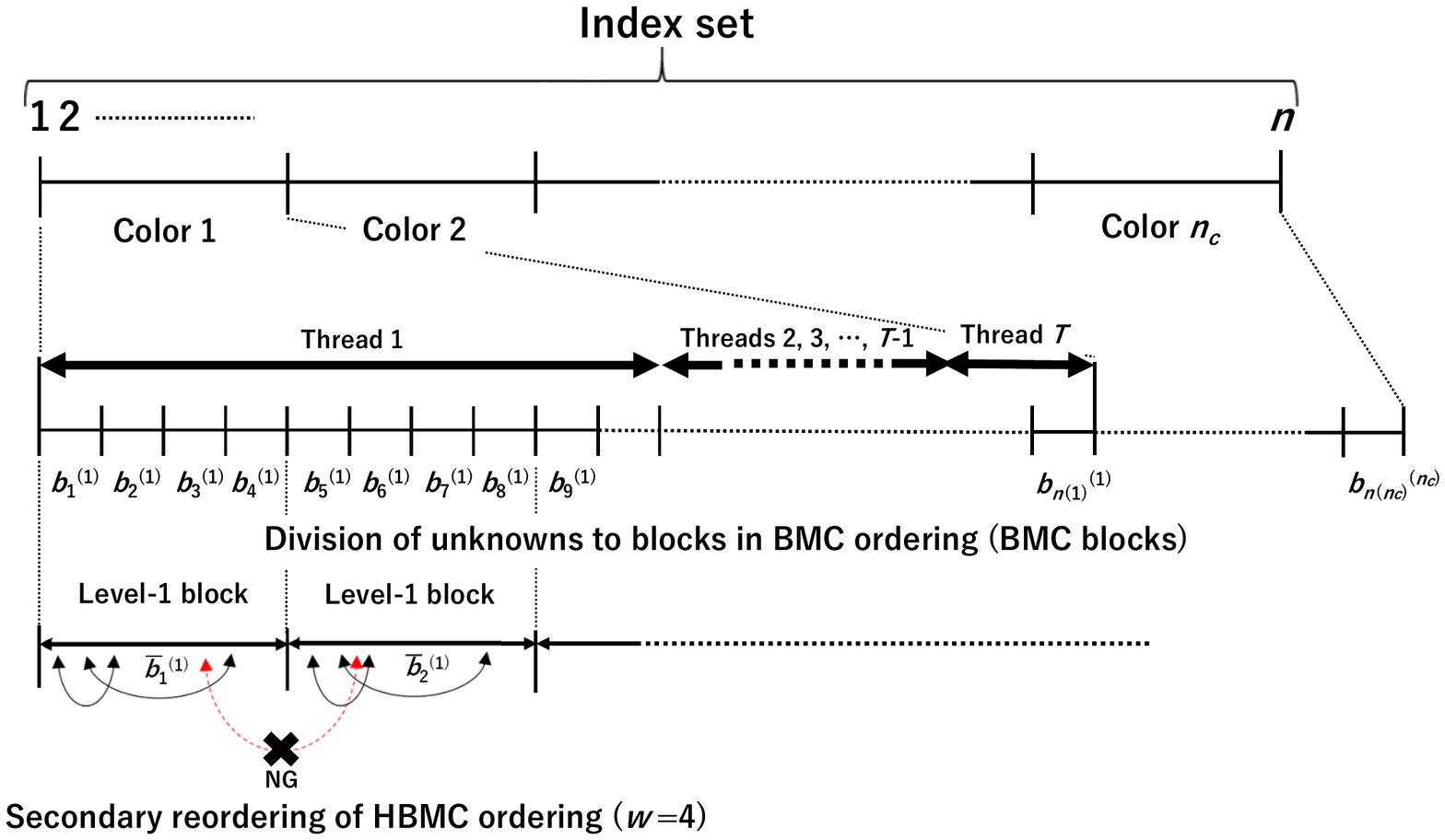} 
\caption{Blocks of BMC and secondary reordering for HBMC}
\label{block-bmc}
\end{figure}

In the present paper, the block size and $k$-th block in color $c$ are denoted by $b_{s}$ and $b_{k}^{(c)}$, respectively.
In BMC, each unknown (or its index) is assigned to a certain block, as shown in Fig. \ref{block-bmc}, where $n(c)$ is the number of blocks in color $c$.

\subsection{Hierarchical block multi-color ordering (HBMC)}
In the proposed HBMC, a new (hierarchical) block structure is introduced. First, we define a level-1 block (or multithreaded block) as follows. The block consists of $w$ consecutive blocks of BMC in each color.
When the $k'$-th level-1 block in color $c$ is written as $\bar{b}_{k'}^{(c)}$,
\beq
\bar{b}_{k'}^{(c)} = \bigcup_{k=k_{s}+1}^{ks+w} b_{k}^{(c)},
\eeq{simd-block}
where $k_{s}=(k'-1) \times w$.
We note that parameter $w$ is determined by the length of the SIMD vector instruction (SIMD width) of the targeted processor. It is typically 4 or 8, and will be larger in the future.

\begin{figure}[tbp]
\centering
\includegraphics[scale=0.7,clip, bb= 10 65 435 320]{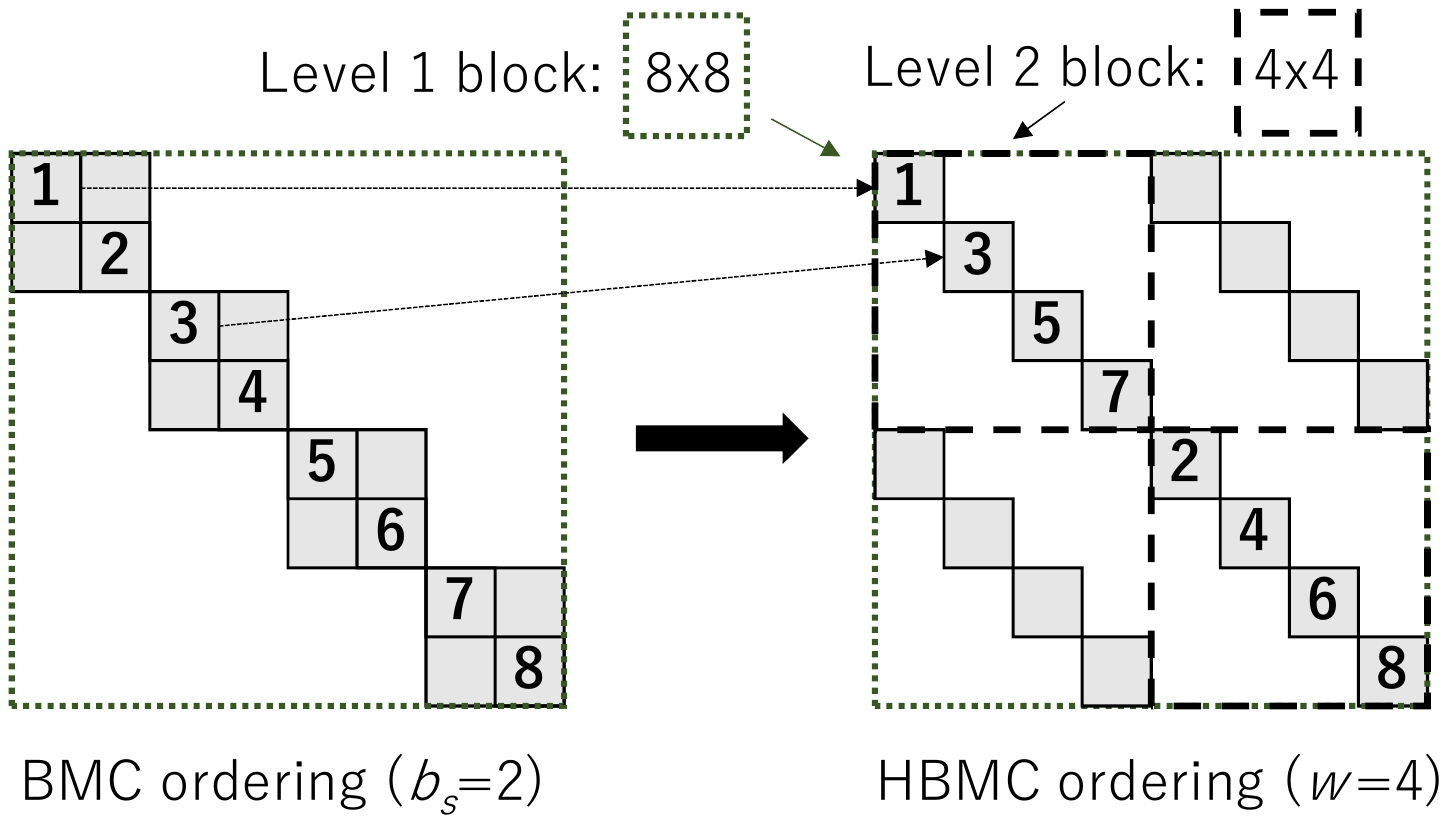} 
\caption{Secondary reordering in a level-1 block}
\label{level1}
\end{figure}

\begin{figure}[tbp]
\centering
\includegraphics[scale=0.65,clip, bb= 105 165 550 515]{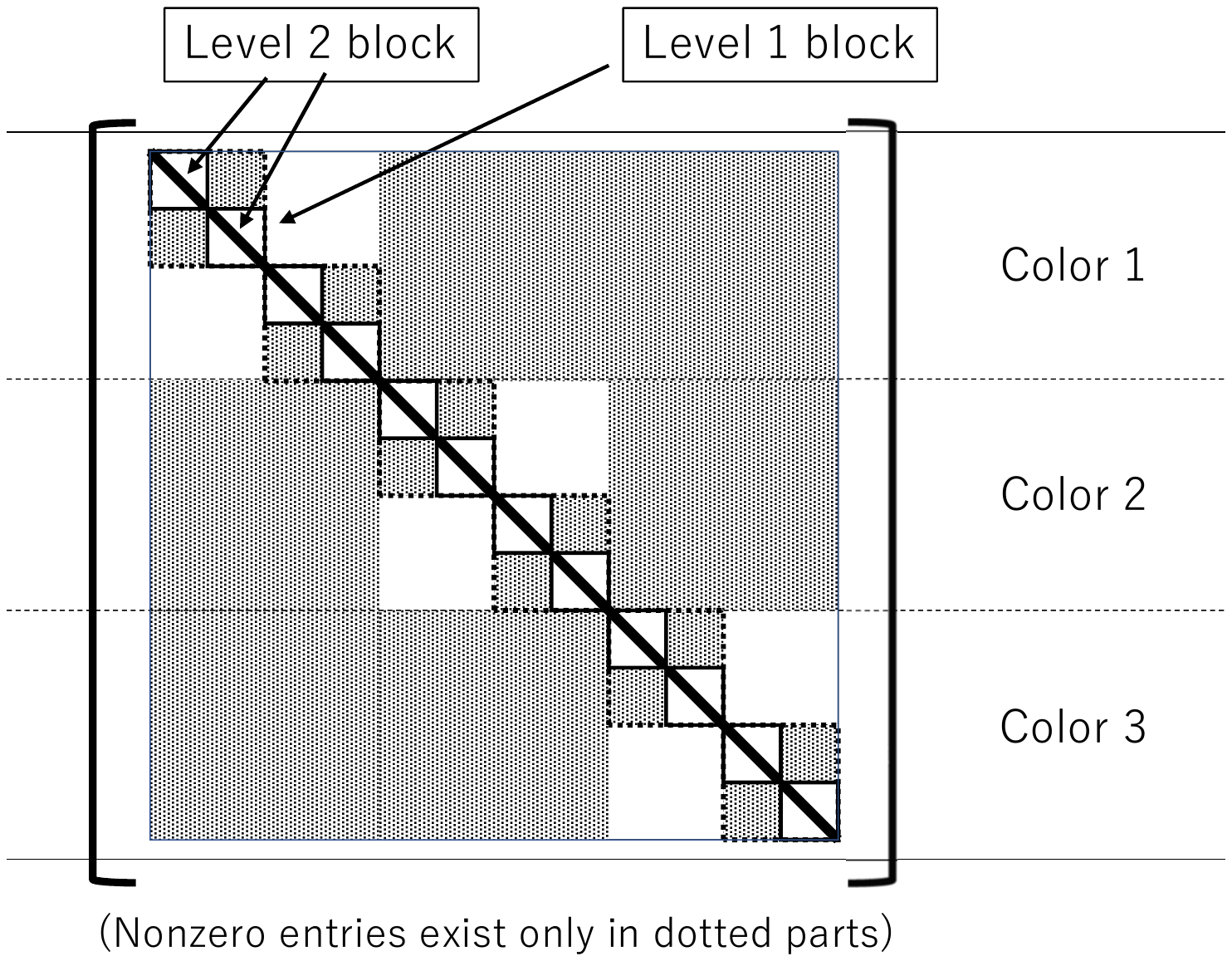}
\caption{Coefficient matrix based on HBMC} 
\label{HBMC-1}
\end{figure}

In our technique, secondary reordering is performed on each level-1 block as shown in Fig. \ref{block-bmc}.
Without loss of generality, we describe the reordering process for a level-1 block, that is, the blocks from $b_{k_{s}+1}^{(c)}$ to
 $b_{k_{s}+w}^{(c)}$ of BMC.
In the first step, we pick up the first (top) unknown of each block, $b_{k_{s}+1}^{(c)}$, $b_{k_{s}+2}^{(c)}, \ldots, b_{k_{s}+w}^{(c)}$, and order the picked unknowns. These $w$ unknowns are mutually independent because the blocks in each color are independent in BMC. In the next step, we pick up the second unknown of each block, which are mutually independent, and order them after the unknowns previously ordered. We repeat the process until no unknowns remain. In total, the pick-up process is performed $b_s$ times. 
%
%
Figure \ref{level1} shows the secondary reordering process in the first level-1 block when $b_{s}=2$ and $w=4$, where each unknown is associated with the diagonal element of the coefficient matrix. In the figure, the colored elements represent nonzero elements.
After the reordering process is complete, we encounter the second-level block structure in the reordered coefficient matrix, which is given by the $w$ $\times$ $w$ (small) diagonal matrices. The level-2 block structure is used for SIMD vectorization of the substitutions.
Figure \ref{HBMC-1} shows the matrix form of the coefficient matrix based on HBMC.

\subsubsection{Equivalence between BMC and HBMC}
We prove that HBMC is equivalent to BMC; that is, the convergence rates of the linear solvers based on the two orderings are the same. 
Because the secondary reordering for HBMC is {\it locally} performed in each level-1 block,   
the order between two unknowns that belong to two different level-1 blocks are preserved in the final order.
Consequently, it holds that 
\begin{eqnarray}
\forall i_{1} \in \bar{b}_{k_{1}}^{(c_{1})}, i_{2} \in \bar{b}_{k_{2}}^{(c_{2})} \ {\rm such \ that} \ c_{1} \neq c_{2} \, \vee \, k_{1} \neq k_{2}, \nonumber \\
\mbox{sgn} (i_{1}-i_{2}) = \mbox{sgn} (\pi (i_{1})-\pi(i_{2})).  
\label{HBMC1}
\end{eqnarray}

From (\ref{HBMC1}), if the {\it local} ordering subgraphs of BMC and HBMC that correspond to each level-1 block are identical, then the two orderings are equivalent.
Next, we examine the reordering process in a level-1 block. 
In the secondary reordering process of HBMC, the order of unknowns that belong to different BMC blocks changes. However, the reordering process for these unknowns does not affect the ordering graph, that is,  the convergence.
In BMC, the unknowns that belong to two different blocks in the same color have no data relationship with one another; that is, there are no edges between them in the ordering graph. Therefore, even if we change the order of unknowns
that belong to different BMC blocks, this does not affect the ordering graph.
Consequently, we now pay attention to the influence of reordering inside a BMC block. When we analyze the above picking process, we can confirm that the order of the unknowns that belong to the same BMC block is preserved in the final order.
In each block, we pick up the first unknown, and then the second, and continue this process. Therefore, the order does not change for these unknowns:
\beq
\forall i_{1}, i_{2} \in b_{k}^{(c)} \ \mbox{sgn} (i_{1}-i_{2}) = \mbox{sgn} (\pi (i_{1})-\pi(i_{2})). 
\eeq{reorder1}
When we consider the mutual independence among the BMC blocks in each color, (\ref{HBMC1}) and (\ref{reorder1}), we can demonstrate that secondary reordering $\pi$ does not change the form of the ordering graph. This proves that HBMC is equivalent to BMC.


\begin{figure*}[tbp]
\centering
\begin{tabular}{cc}
\includegraphics[scale=0.52,clip, bb= 0 60 388 495]{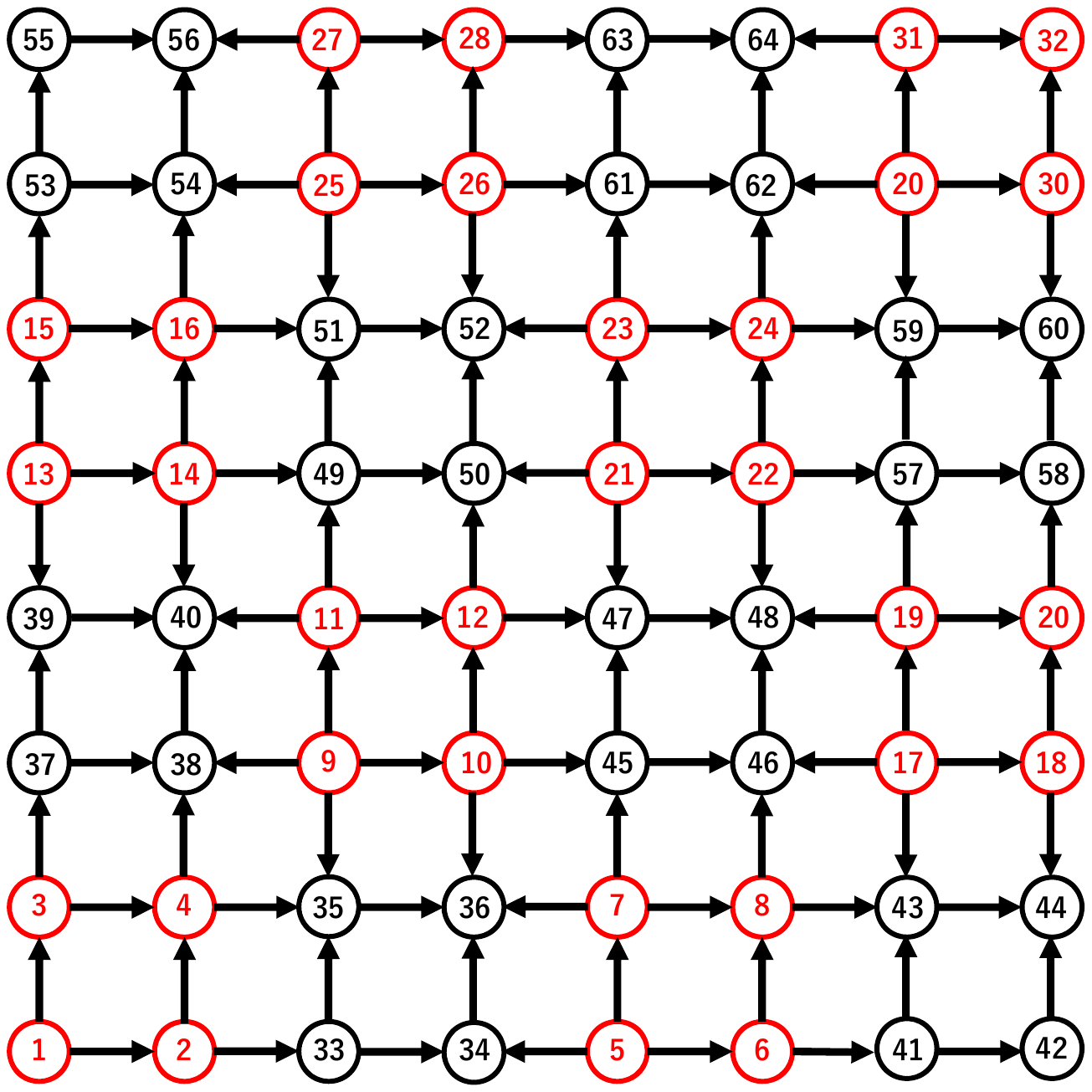} &
\includegraphics[scale=0.52,clip, bb= 0 60 388 495]{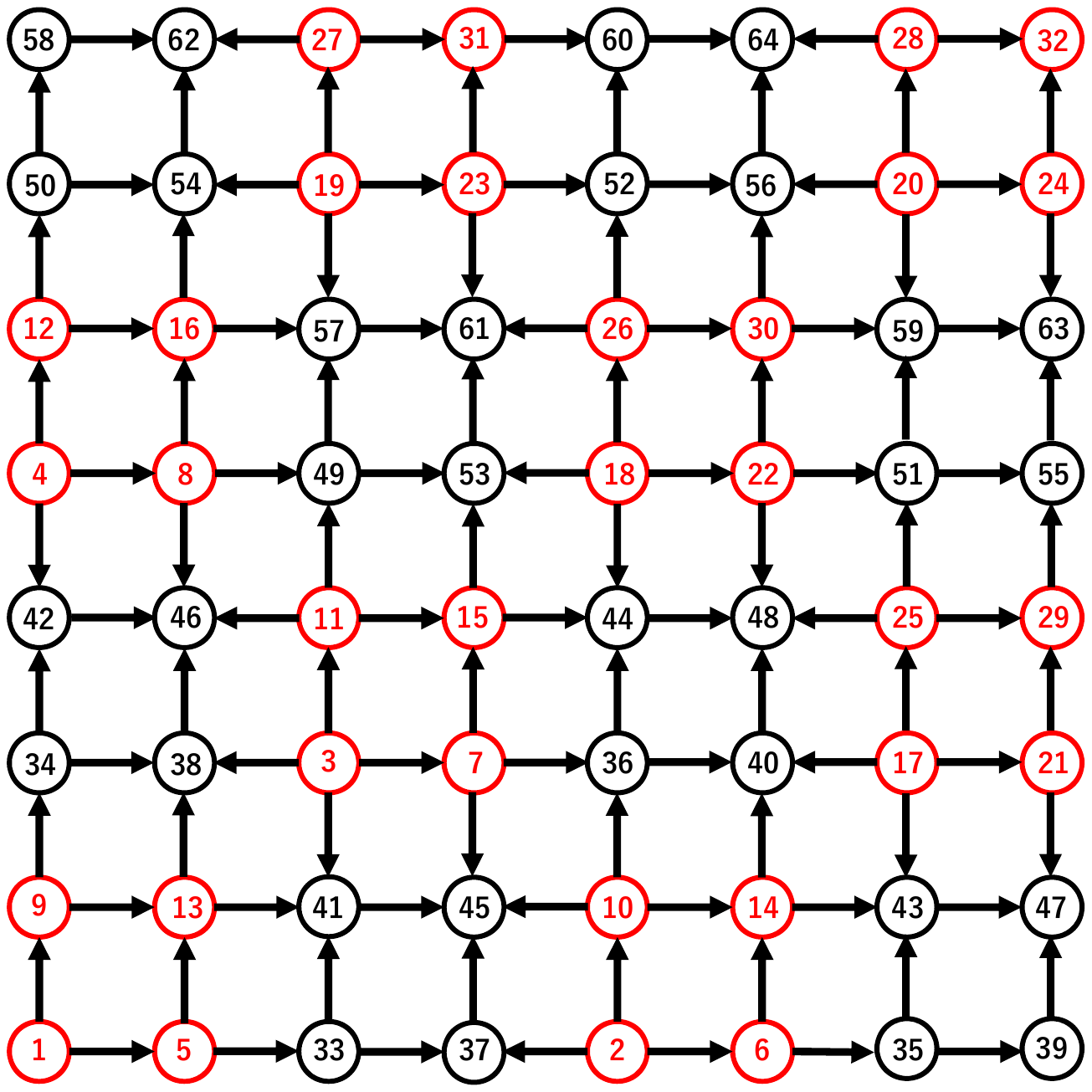} \\
(a) BMC ($b_{s}=4$, $n_{c}=2$) & (b) HBMC  ($w=4$, $b_{s}=4$, $n_{c}=2$) \\
& \\
\multicolumn{2}{c}{\includegraphics[scale=0.5,clip, bb= 8 70 458 510]{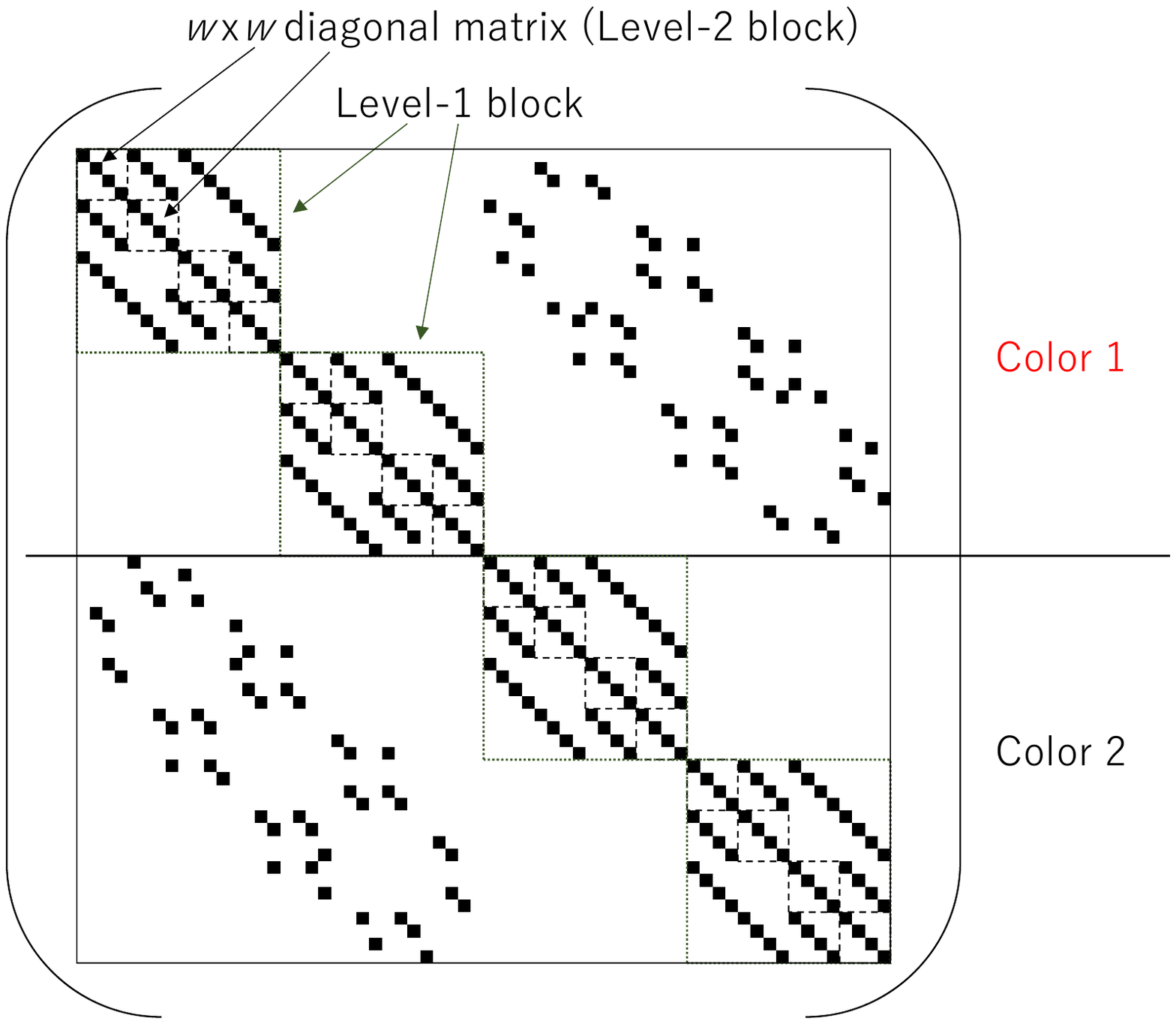} }\\
\multicolumn{2}{c}{(c) Coefficient matrix based on HBMC} \\
\end{tabular}
\caption{Ordering graphs in a five-point stencil problem and coefficient matrix based on HBMC}
\label{5point-bmc-hbmc}
\end{figure*}

As an example that shows the relationship between BMC and HBMC, Fig. \ref{5point-bmc-hbmc} demonstrates the ordering of nodes (unknowns) in a five-point finite difference analysis. Figures \ref{5point-bmc-hbmc} (a) and (b) show that BMC and HBMC have identical ordering graphs. Consequently, the two orderings are equivalent in terms of convergence. Figure \ref{5point-bmc-hbmc} (c) shows the coefficient matrix based on HBMC, which involves the hierarchical block structures.

\subsection{Parallelization and vectorization of forward and backward substitutions}
Corresponding to the colors of the unknowns, solution vector $\bar{\xxx}$ and coefficient matrix $\bar{\AAA}$ are split as 
\beq
\bar{\xxx}= \left(   \begin{array}{c}
\bar{\xxx}_{1} \\
\bar{\xxx}_{2} \\
\vdots \\
\bar{\xxx}_{n_{c}} \\
\end{array}
\right),
\eeq{x}
and
%
\beq
\bar{\AAA}= \left(   \begin{array}{cccc}
\bar{\CCC}_{1,1} & \bar{\CCC}_{1,2} & \ldots & \bar{\CCC}_{1,n_{c}}\\
\bar{\CCC}_{2,1} & \bar{\CCC}_{2,2} & \ldots & \bar{\CCC}_{2,n_{c}}\\
\vdots & \vdots & \ddots & \vdots\\
\bar{\CCC}_{n_{c},1} & \bar{\CCC}_{n_{c},2} & \ldots & \bar{\CCC}_{n_{c},n_{c}} \\
\end{array}
\right),
\eeq{ax}
where $\bar{\xxx}_{c}$ corresponds to the unknowns with color $c$, and $\bar{\CCC}_{c,d}$ represents the relationship between $\bar{\xxx}_{c}$ and $\bar{\xxx}_{d}$. 
Hereafter, we assume that the size of $\bar{\xxx}_{c}$ is a multiple of $b_{s}w$. In the analysis program, the assumption is satisfied using some dummy unknowns. 
Let the number of level-1 blocks assigned to color $c$ be denoted by $\bar{n}(c)$, then diagonal block $\bar{\CCC}_{c,c}$ of $\bar{\AAA}$ is given by the following block diagonal matrix:
\beq
\bar{\CCC}_{c,c} = \left(   \begin{array}{cccc}
\bar{\BBB}_{1}^{(c)} &  & & \0 \\
 & \bar{\BBB}_{2}^{(c)} &  &  \\
  &  & \ddots & \\
\0 &  & & \bar{\BBB}_{\bar{n}(c)}^{(c)} \\
\end{array}
\right),
\eeq{BAMC1}
where $\bar{\BBB}_{k}^{(c)}$ is the $b_{s}w$ $\times$ $b_{s}w$ matrix that corresponds to the unknowns in the $k$-th level-1 block with color $c$, which we denote by $\bar{b}_{k}^{(c)}$.
Moreover, matrix $\bar{\BBB}_{k}^{(c)}$ is written as
\beq
\bar{\BBB}_{k}^{(c)} = \left(   \begin{array}{cccc}
\bar{\DDD}_{1}^{(k, c)} & \bar{\EEE}_{1,2}^{(k,c)} & \ldots & \bar{\EEE}_{1,b_{s}}^{(k,c)} \\
 \bar{\EEE}_{2,1}^{(k,c)} & \bar{\DDD}_{2}^{(k, c)} & \ldots &  \bar{\EEE}_{2,b_{s}}^{(k,c)} \\
 \vdots & \vdots & \ddots & \\
  \bar{\EEE}_{b_{s},1}^{(k,c)}& \bar{\EEE}_{b_{s},2}^{(k,c)} & \ldots & \bar{\DDD}_{b_{s}}^{(k, c)} \\
\end{array}
\right),
\eeq{BAMC-level-2}
where $\bar{\DDD}_{l}^{(k, c)}, (l=1, 2, \ldots, b_{s})$ are $w$ $\times$ $w$ diagonal matrices.

The forward substitution included in ILU(0)/IC(0) preconditioners or GS and SOR methods uses a lower triangular matrix with the same nonzero element pattern as the lower triangular part of $\bar{\AAA}$. From (\ref{ax}) and (\ref{BAMC1}), lower triangular matrix $\bar{\LLL}$ is written as
\beq
\bar{\LLL}= \left(   \begin{array}{cccc}
\bar{\LLL}_{1,1} &  & & \\
\bar{\LLL}_{2,1} & \bar{\LLL}_{2,2} &  & \0 \\
\vdots & \ddots & \ddots & \\
\bar{\LLL}_{n_{c},1} & \bar{\LLL}_{n_{c},2} & \ldots & \bar{\LLL}_{n_{c},n_{c}} \\
\end{array}
\right), 
\eeq{lx}
and diagonal block $\bar{\LLL}_{c,c}$ is given by 
\beq
\bar{\LLL}_{c,c} = \left(   \begin{array}{cccc}
\bar{\LLL}_{1}^{(c)} &  & & \0 \\
 & \bar{\LLL}_{2}^{(c)} &  &  \\
  &  & \ddots & \\
\0 &  & & \bar{\LLL}_{\bar{n}(c)}^{(c)} \\
\end{array}
\right),
\eeq{BAMC2}
where $\bar{\LLL}_{k}^{(c)}$ is the $b_{s}w$ $\times$ $b_{s}w$ lower triangular matrix that corresponds to block $\bar{b}_{k}^{(c)}$.
The forward substitution for the reordered linear system is given by
\beq
\bar{\LLL} \bar{\yyy} = \bar{\rrr},
\eeq{for1}
where $\bar{\rrr}$ is the residual vector in the case of ILU (IC) preconditioning. 
Let $\bar{\yyy}_{c}$ and $\bar{\rrr}_{c}$ represent, respectively, the segments of $\bar{\yyy}$ and $\bar{\rrr}$ that correspond to color $c$, and $\bar{\yyy}_{k}^{(c)}$ and $\bar{\rrr}_{k}^{(c)}$ represent the subsegments in the segments that correspond to block $\bar{b}_{k}^{(c)}$; that is,
\beq
\bar{\yyy}_{c}= \left(   \begin{array}{c}
\bar{\yyy}_{1}^{(c)} \\
\bar{\yyy}_{2}^{(c)} \\
\vdots \\
\bar{\yyy}_{\bar{n}(c)}^{(c)} \\
\end{array}
\right), \ \mbox{and} \ \ 
\bar{\rrr}_{c}= \left(   \begin{array}{c}
\bar{\rrr}_{1}^{(c)} \\
\bar{\rrr}_{2}^{(c)} \\
\vdots \\
\bar{\rrr}_{\bar{n}(c)}^{(c)} \\
\end{array}
\right).
\eeq{y1}
%
Then, from (\ref{lx}) and (\ref{for1}), the forward substitution for $\bar{\yyy}_{c}$ is given by 
\beq
\bar{\yyy}_{c} = \bar{\LLL}_{c,c}^{-1} \bar{\qqq}_{c},
\eeq{for2}
where
\beq
\bar{\qqq}_{c} = \bar{\rrr}_{c} -  \sum_{d=1}^{c-1} \bar{\LLL}_{c,d} \bar{\yyy}_{d}.
\eeq{for3}
Because vector segments $\bar{\yyy}_{d} (d=1, \ldots, c-1)$ are computed prior to the substitution (\ref{for2}) and shared among all threads, $\bar{\qqq}_{c}$ is a given vector in (\ref{for2}). When the segment of $\bar{\qqq}_{c}$ that corresponds to block $\bar{b}_{k}^{(c)}$ is denoted by $\bar{\qqq}_{k}^{(c)}$ as in (\ref{y1}), from (\ref{BAMC2}), the forward substitution (\ref{for2}) is expressed as $\bar{n}(c)$ independent steps:
\beq
\bar{\yyy}_{k}^{(c)}=(\bar{\LLL}_{k}^{(c)})^{-1} \bar{\qqq}_{k}^{(c)} \ (k=1, \ldots, \bar{n}(c)).
\eeq{BAMC3}
Consequently, the forward substitution (\ref{for2}) for color $c$ can be multithreaded with the degree of parallelism given by the number of level-1 blocks of each color, which is approximately $n/(n_{c} \cdot b_{s} \cdot w)$. Each thread processes one or more level-1 blocks in parallel. 

Next, we explain how to vectorize each step of  (\ref{BAMC3}).
We consider the procedure for the $k$-th level-1 block of color $c$: $\bar{\yyy}_{k}^{(c)}=(\bar{\LLL}_{k}^{(c)})^{-1} \bar{\qqq}_{k}^{(c)}$.
From (\ref{BAMC-level-2}), lower triangular matrix $\bar{\LLL}_{k}^{(c)}$ is written as
\beq
\bar{\LLL}_{k}^{(c)} = \left(   \begin{array}{cccc}
\tilde{\DDD}_{1}^{(k, c)} &  &  & \\
 \bar{\LLL}_{2,1}^{(k,c)} & \tilde{\DDD}_{2}^{(k, c)} &  & \0 \\
 \vdots & \ddots & \ddots & \\
  \bar{\LLL}_{b_{s},1}^{(k,c)}& \bar{\LLL}_{b_{s},2}^{(k,c)} & \ldots & \tilde{\DDD}_{b_{s}}^{(k, c)} \\
\end{array}
\right),
\eeq{BAMC-level-3}
where $\tilde{\DDD}_{l}^{(k, c)}$ are diagonal matrices. 
We split $\bar{\yyy}_{k}^{(c)}$ and $\bar{\qqq}_{k}^{(c)}$ into $b_{s}$ segments, each of size $w$.
Let $\bar{\yyy}_{l}^{(k, c)}$ and $\bar{\qqq}_{l}^{(k, c)}$ represent the segments that correspond to the level-2 block of the $k$-th level-1 block of color $c$; that is,
\beq
\bar{\yyy}_{k}^{(c)}= \left(   \begin{array}{c}
\bar{\yyy}_{1}^{(k, c)} \\
\bar{\yyy}_{2}^{(k, c)} \\
\vdots \\
\bar{\yyy}_{b_{s}}^{(k, c)} \\
\end{array}
\right), \ \mbox{and} \ \ 
\bar{\qqq}_{k}^{(c)}= \left(   \begin{array}{c}
\bar{\qqq}_{1}^{(k, c)} \\
\bar{\qqq}_{2}^{(k, c)} \\
\vdots \\
\bar{\qqq}_{b_{s}}^{(k, c)} \\
\end{array}
\right).
\eeq{level2-y1}
Then, from (\ref{BAMC-level-3}), the forward substitution for level-1 block $\bar{b}_{k}^{(c)}$ is given by the following $b_{s}$ sequential steps:
\beq
\bar{\yyy}_{l}^{(k, c)} = (\tilde{\DDD}_{l}^{(k, c)})^{-1} \bar{\ttt}_{l}^{(k, c)}, (l=1,2,\ldots,b_{s}), 
\eeq{y2-2}
where
\beq
\bar{\ttt}_{l}^{(k, c)} =\bar{\qqq}_{l}^{(k, c)} - \sum_{m=1}^{l-1}\bar{\LLL}_{l,m}^{(k,c)} \bar{\yyy}_{m}^{(k, c)}. 
\eeq{y2-3}
In the $l$-th step of (\ref{y2-2}), because $\tilde{\DDD}_{l}^{(k, c)}$ is a diagonal matrix and each element of $\bar{\ttt}_{l}^{(k, c)}$ can be calculated in parallel, the step consists of $w$ independent steps that can be efficiently vectorized. In other words, the $l$-th step of (\ref{y2-2}) consists of a simple matrix vector multiplication and element-wise vector updates that are directly vectorized.

The backward substitution is parallelized (multithreaded) and vectorized in a similar manner, although it is performed inversely from color $n_{c}$ to 1.

\subsection{Implementation of HBMC}
\subsubsection{Reordering process}
In this section, we discuss the reordering process.
In the technique, any type of algorithm (heuristic) for an implementation of BMC can be used.
In the application of BMC, we set the number of BMC blocks assigned to each thread as a multiple of $w$, except for one of the threads (typically the last-numbered thread). In this circumstance, the application of HBMC, that is, the secondary reordering from BMC, is performed in each thread. Therefore, the reordering process is fully multithreaded.

\subsubsection{Storage format}
In the implementation of the sparse triangular solver, a storage format for sparse matrices~\cite{templates} is typically used.
For example, the factorization matrices in an IC/ILU preconditioned solver are stored in memory using such a format.
Although there are several standard formats, 
the sliced ELL (SELL) format~\cite{sell-1} is the most efficient for exploiting the benefit of SIMD instructions, and we used it in our implementation. 
In the SELL format, the slice size is an important parameter. In HBMC, we naturally set the slice size as $w$ because the forward and backward substitutions are vectorized every $w$ rows. This leads to the natural introduction of the concept of the SELL-C-$\sigma$ format~\cite{sell-2} to the analysis, which is a sophisticated version of SELL.

\subsubsection{Multithreaded and vectorized substitutions}
The program for each forward and backward substitution consists of nested loops. The outer-most loop is for the color. After the computations for each color, thread synchronization is required. Therefore, the number of synchronizations in each substitution is given by $n_{c}-1$, which is the same as BMC and the standard multi-color ordering (MC).  The second loop is for level-1 blocks. Because the level-1 blocks in each color are mutually independent, each thread processes single or multiple level-1 blocks in parallel.
In each level-1 block, the substitution can be split into $b_{s}$ steps, each of which is vectorized with a SIMD width of $w$.
For the vectorized substitution, we used the OpenMP SIMD directive or the Intel intrinsic functions for SIMD instructions. Figure \ref{C-vec} shows a sample C program code for multithreaded and vectorized forward substitution using OpenMP and Intel AVX-512 intrinsic functions.

Additionally, we discuss the special nonzero pattern that appears in $\bar{\LLL}_{k}^{(c)}$ that corresponds to the level-1 block. In the matrix, all nonzero elements lay on $2b_{s}-1$ diagonal lines. Although we can consider using a hybrid storage format that exploits this special structure, it does not typically result in better performance because of the additional cost of processing the diagonal block and other off-diagonal elements separately. We confirmed this in some preliminary tests.

Finally, we discuss the data access locality. The access pattern for the vector elements in HBMC is different from that in BMC. Therefore, the data access locality can be different between two orderings. However, because the secondary reordering for HBMC is performed inside a level-1 block, the data access locality barely changes; at least, from the viewpoint of the last-level cache, both orderings are considered to be similar.

\begin{figure}[tb]
\begin{center}
\begin{tabular}{|l|}\hline
\verb|for (c = 0; c < nc; c++){|\\
\verb|  #pragma omp for private(p, j, num, t, index, \|\\
\verb|                     mtmp, mval, pos, mb, mdiag)|\\
\verb|  for ( k = lev1b[c]; k < lev1b[c+1]; k++ ){|\\
\verb|    num = k * 8 * bs;|\\
\verb|    index = mat.offset[k];|\\
\verb|    j = lev2b[k];|\\
\verb|    for ( p = j; p < j + bs; p++ ){|\\
\verb|      mtmp = _mm512_load_pd( &r[num] );|\\
\verb|      for ( t = 0; t < mat.slen.lev2b[p]; t++ ){|\\
\verb|        mval = _mm512_load_pd( &val[index] );|\\
\verb|        pos = _mm512_load_epi32( &col[index] );|\\
\verb|        mb = _mm512_i32logather_pd( pos, z, 8 );|\\
\verb|        mtmp = _mm512_sub_pd( mtmp, \ |\\
\verb|                   _mm512_mul_pd(mval,mb) );|\\
\verb|        index += 8;|\\
\verb|      }|\\
\verb|      mdiag = _mm512_load_pd( &diaginv[num] );|\\
\verb|      mtmp = _mm512_mul_pd( mtmp, mdiag );|\\
\verb|      _mm512_store_pd( &z[num], mtmp );|\\
\verb|      num = num + 8;|\\
\verb|    }|\\
\verb|  }|\\
\verb|}|\\ \hline
\end{tabular}
\caption{C program code of multithreaded and vectorized forward substitution with OpenMP directives and Intel AVX-512 intrinsic functions}
\label{C-vec}
\end{center}
\end{figure}

\begin{table*}[tbp]
	\centering
	\caption{Information about the test environments}
	\label{cpuInfo}
	\begin{tabular}{|c|c|c|c|}
	\hline
    System   & Cray XC40 & Cray CS400 (2820XT)  & Fujitsu CX2550 (M4)  \\ 	\hline
    \multirow{2}{*}{Processor type} & Intel Xeon Phi & Intel Xeon  & Intel Xeon  \\ 
                          &    (7250, KNL)      & (E5-2695 v4, Broadwell)   &   (Gold6148, Skylake)     \\ \hline    
   \# processors / node & 1 & 2  & 2 \\ \hline
    \# cores / processor  & 68 & 18 & 20 \\ \hline
    Clock (GHz)  & 1.4 & 2.1 & 2.4 \\ \hline
    Compiler &  icc 18.0.3 &  icc 17.0.6      & icc 17.0.6 \\ \hline
    \multirow{3}{*}{Compile options} & -qopenmp -ipo -O3 & -mcmodel=medium & -mcmodel=medium          \\
                          &   -xMIC-AVX512  &  -shared-intel -qopenmp  & -shared-intel -qopenmp          \\
                          &                            &  -xHost    -ipo -O3                          &  -xCORE-AVX512 -ipo -O3   \\ \hline
    \end{tabular}
\end{table*}

\section{Numerical Results}
\subsection{Computers and Test Problems}
We conducted five numerical tests on three types of computational nodes to evaluate the proposed reordering technique in the context of the ICCG method: the computational nodes were Cray XC40, Cray CS400 (2820XT), and Fujitsu CX2550 (M4). The two Cray systems are operated by Academic Center for Computing and Media Studies, Kyoto Univ., whereas the Fujitsu system is at Information Initiative Center, Hokkaido Univ. Table \ref{cpuInfo} lists information about the computational node and compiler used. In the numerical tests, we used all cores of the computational node for execution.

The program code was written in C and OpenMP for the thread parallelization. 
For vectorization, we used the intrinsic functions of the Intel Advanced Vector Extensions.
The AVX2 (256 bits SIMD) instruction set was used for the Xeon (Broadwell) processor, whereas the AVX-512 (512 bits SIMD) instruction set was used for the Xeon Phi (KNL) and Xeon (Skylake) processors. 
Although we also developed a vectorized program using the OpenMP SIMD directive, its performance was slightly worse than the version with the intrinsic function in most of the test cases.
Thus, in this paper, we only report the results from using the intrinsic function.
%

For the test problems, we used a linear system that arises from finite element electromagnetic field analysis and four linear systems picked up from the SuiteSparse Matrix Collection.
We selected symmetric positive-definite matrices that are mainly derived from computational science or engineering problems, and have a relatively large 
dimension compared with other matrices in the collection.

In the electromagnetic field analysis test, the linear system that arises from the finite element discretization of the IEEJ standard benchmark model~\cite{ieej} was used. The basic equation for the problem is given as
\beq
\nabla \times (\nu \nabla \times \AAA_{m}) = \JJJ_{0},
\eeq{ieej}
where $\AAA_{m}$, $\nu$, and $\JJJ_{0}$ are the magnetic vector potential, magnetic reluctivity, and excitation current density,
respectively. The analysis solved (\ref{ieej}) using a
finite edge-element method with hexahedral elements.
Applying the Galerkin method to (\ref{ieej}), we obtained a linear
system of equations for the test problem.
 The resulting linear system was solved using the shifted ICCG method, with the shift parameter given as 0.3.
The name of the dataset of the linear system is denoted by Ieej.
Table \ref{matInfo} lists the matrix information for the test problems.

In this paper, we report the performance comparison of four multithreaded ICCG solvers.
The solver denoted by ``MC'' is based on the multi-color ordering which is the most popular parallel ordering method.
The solver ``BMC'' is the solver based on the block multi-color ordering method.
The solvers ``HBMC (crs\_spmv)'' and ``HBMC (sell\_spmv)'' are based on the proposed HBMC, where the former solver used compressed row storage (CRS) format~\cite{templates} for the implementation of sparse matrix vector multiplication (SpMV) and the latter used the SELL format. In MC and BMC, the CRS format was used. 


For the blocking method in BMC and HBMC, we used the simplest one among the heuristics introduced in \cite{IPDPS2012}, in which the unknown with the minimal number is picked up for the newly generated block. For the coloring of nodes or blocks, the greedy algorithm was used for all the solvers. The convergence criterion was set as the relative residual norm (2-norm)
being less than $10^{-7}$.

%
%

\begin{table}[tbp]
	\centering
	\caption{Matrix information for the test problems}
	\label{matInfo}
	\begin{tabular}{|c|c|c|c|}
	\hline
	Data set & Problem type & Dimension & \# nonzero \\
	\hline
	\multirow{2}{*}{Thermal2}       & Thermal                      & \multirow{2}{*}{1,228,045} & \multirow{2}{*}{8,580,313}  \\ 
	                   &  problem                     &  &   \\ \hline
	Parabolic\_fem & CFD & 525,825   & 3,674,625  \\ \hline
	G3\_circuit    & Circuit problem           & 1,585,478 & 7,660,826  \\ \hline
    \multirow{2}{*}{Audikw\_1}      & Structural            & \multirow{2}{*}{943,695}   & \multirow{2}{*}{77,651,847} \\
                       &  problem & & \\ \hline
 \multirow{2}{*}{Ieej}     & Eddy current      & \multirow{2}{*}{1,011,920} & \multirow{2}{*}{31,468,056} \\
             & analysis & & \\ \hline
	\end{tabular}
\end{table}

\begin{table}[tbp]
	\centering
	\caption{Comparison of the number of iterations}
	\label{number-ite}
	\begin{tabular}{|c|c|c|c|}
        \hline
       Dataset$\backslash$method & MC & BMC & HBMC \\
        \hline
        Thermal2  & 2283     & 2129  & 2129 \\
        \hline
        Parabolic\_fem & 1145 & 1052  & 1052 \\
        \hline
        G3\_circuit  & 1521  & 1227 & 1227 \\
        \hline
        Audikw\_1 & 1728   & 1714 & 1715 \\
        \hline
        Ieej  & 580  & 446    & 446  \\
        \hline
	\end{tabular}
\end{table}

\subsection{Numerical results}
\subsubsection{Equivalence of orderings in convergence and use of SIMD instructions}
\begin{figure}[tbp]
\centering
\includegraphics[scale=0.7,clip, bb= 20 45 490 350]{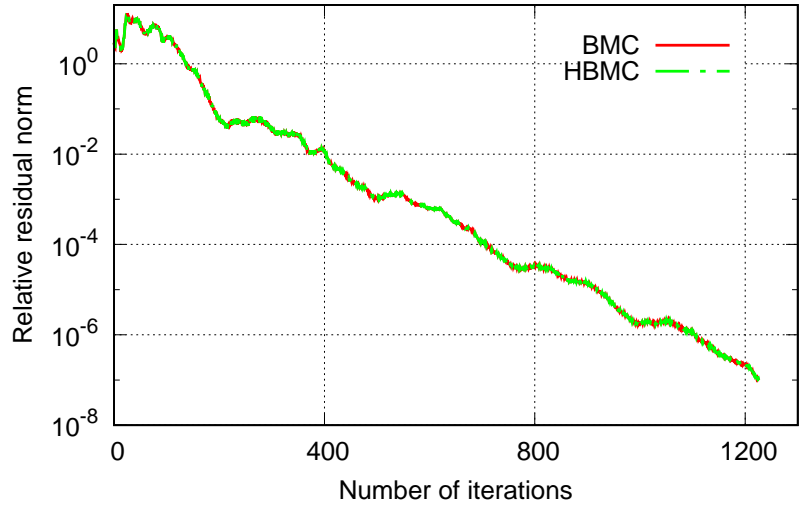} \\
(a) G3\_circuit test \\
\includegraphics[scale=0.7,clip, bb= 20 45 490 350]{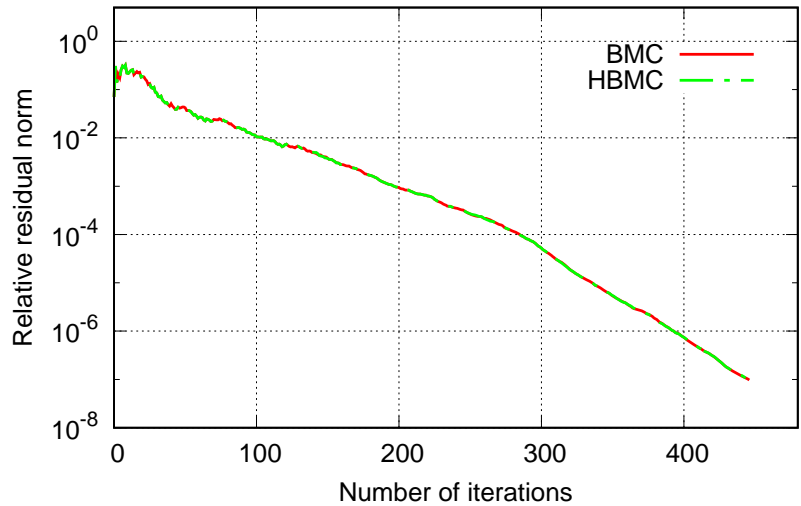} \\
(b) Ieej test
\caption{Convergence behavior of BMC and HBMC}
\label{conv-behavior}
\end{figure}

First, we examine the equivalence of BMC and HBMC in terms of convergence.
Table \ref{number-ite} lists the number of iterations of the solvers tested on Cray XC40, where the block size of BMC and HBMC was set to 32. Equivalence was confirmed by the numerical results. Moreover, to examine the entire solution process, Figure \ref{conv-behavior} shows the convergence behaviors of BMC and HBMC in the G3\_circuit and Ieej tests. In the figure, the two lines of the relative residual norms overlap, which indicates that the solvers had an equivalent solution process. The equivalence of convergence was also confirmed in all test cases (five datasets $\times$ three block sizes $\times$ three computational nodes). Furthermore, Table \ref{number-ite} shows the advantage in terms of convergence of BMC over MC, which coincides with the results reported in \cite{IPDPS2012}.

Next, we checked the use of SIMD instructions in the solver using the Intel Vtune Amplifier (application performance snapshot) in the G3\_circuit test conducted on Fujitsu CX2550. The snapshot showed that the percentage of all packed floating point instructions in the solver based on HBMC (sell\_spmv) reached 99.7\%, although that in the solver using BMC was 12.7\%.


\subsubsection{Performance comparison}
Table \ref{Results} (a) shows the performance comparison of four solvers in the numerical tests on Cray XC40.
In the tests, HBMC attained the best performance for all datasets, except Audikw\_1.
In the Thermal2 and G3\_circuit tests, HBMC (sell\_spmv) was more than two times faster than the standard MC solver.
When HBMC (crs\_spmv) was compared with BMC, it attained better performance in 11 out of 15 cases (five datasets $\times$ three block sizes), which demonstrates the effectiveness of HBMC for the sparse triangular solver.
Moreover, in all test cases, HBMC (sell\_spmv) outperformed HBMC (crs\_spmv), which implies that an efficient use of SIMD instructions was important on the Xeon Phi-based system.

Table \ref{Results} (b) shows the test results on Cray CS400 (Xeon Broadwell).
In the numerical tests, HBMC attained the best performance for all datasets.
When HBMC (crs\_spmv) was compared with BMC, it attained better performance in 13 out of 15 cases, which shows the effectiveness of HBMC.
Table \ref{Results} (b) also indicates that using the SELL format  for the coefficient matrix mostly led to an improvement in solver performance.

Table \ref{Results} (c) shows the test results on Fujitsu CX2550 (Xeon Skylake).
In the numerical tests, HBMC outperformed MC and BMC for four out of five datasets.
For the Audikw\_1 dataset, HBMC did not outperform BMC on Xeon Phi and Skylake, although it was better than BMC on Xeon Broadwell. This result is thought to be
because of the effect of increasing the slice size that was given by the SIMD width, $w$.
In the SELL format, some zero elements were considered as nonzero in the slice.
When there was a significant imbalance among the number of nonzero elements in each row in the slice, the number of elements processed largely increased compared with the implementation with the CRS format.
The Audikw\_1 dataset had this property.
For Audikw\_1, the number of processed elements in SELL increased by 40\% compared with that in CRS, although the increase was 10\% for the G3\_circuit dataset.
For this type of dataset, when the size of the slice increased, the number of processed elements often increased.
The size of the slice was set to 8 for the Xeon Phi and Skylake processors and 4 for the Xeon Broadwell processors.
On the Broadwell processors, the increase in the number of elements when changing CRS to SELL was 28\%, which resulted in better performance for HBMC compared with BMC. In the future, for further acceleration of the solver, we will develop an implementation in which we introduce some remedies for this SELL issue, for example, splitting in two the row that includes an extremely large number of nonzero elements compared with other rows.

\begin{table*}[t]
\centering
\caption{Numerical results: execution time (sec.)}
\label{Results}
(a) Cray XC40 (Intel Xeon Phi)

\vspace{0.3\baselineskip}

\begin{tabular}{|c|c|ccc|ccc|ccc|}
		\hline
            \multirow{2}{*}{Dataset} & \multirow{2}{*}{MC} & \multicolumn{3}{|c}{BMC} & \multicolumn{3}{|c}{HBMC (crs\_spmv)} & \multicolumn{3}{|c|}{HBMC (sell\_spmv)} \\
			& & $b_{s}=8$ & $b_{s}=16$ & $b_{s}=32 $ & $b_{s}=8$ & $b_{s}=16$ & $b_{s}=32 $ & $b_{s}=8$ & $b_{s}=16$ & $b_{s}=32 $ \\ \hline
		Thermal2		& 20.2  & 17.8 & 17.9 & 19.4 & 13.8 & 14.5 & 15.6 & {\bf 9.28} & 9.79 & 11.1 \\
		Parabolic\_fem	& 2.64  & 2.77 & 2.79 & 2.72 & 2.35 & 2.33 & 2.48 & {\bf 1.72} & {\bf 1.72} & 1.79 \\
		G3\_circuit		& 7.98  & 7.97 & 7.72 & 9.91 & 5.16 & 5.27 & 5.49 & {\bf 3.35} & 3.55 & 3.61 \\
		Audikw\_1		& 109.4 & 73.2 & 69.3 & {\bf 63.9} & 72.7 & 73.2 & 73.3 & 68.3 & 68.0 & 69.1 \\
		Ieej			& 4.58  & 5.35 & 5.07 & 4.12 & 4.94 & 5.02 & 4.55 & 3.60 & 3.49 & {\bf 3.18} \\
		\hline
		\end{tabular}

\vspace{0.3\baselineskip}

(b) Cray CS400 (Intel Xeon Broadwell)

\vspace{0.3\baselineskip}

\begin{tabular}{|c|c|ccc|ccc|ccc|}
		\hline
            \multirow{2}{*}{Dataset} & \multirow{2}{*}{MC} & \multicolumn{3}{|c}{BMC} & \multicolumn{3}{|c}{HBMC (crs\_spmv)} & \multicolumn{3}{|c|}{HBMC (sell\_spmv)} \\
			& & $b_{s}=8$ & $b_{s}=16$ & $b_{s}=32 $ & $b_{s}=8$ & $b_{s}=16$ & $b_{s}=32 $ & $b_{s}=8$ & $b_{s}=16$ & $b_{s}=32 $ \\ \hline
		Thermal2		      & 16.3  & 14.8 & 14.8 & 15.3 & {\bf 13.5} & 14.2 & 14.8 & 14.4        & 14.2 & 14.7 \\
		Parabolic\_fem	& 2.91  & 2.69 & 2.69 & 2.53 & 2.55       & 2.54 & 2.51 & 2.41        & 2.33 & {\bf 2.31} \\
		G3\_circuit		& 9.39  & 7.70 & 8.15 & 7.92 & 7.62       & 7.90 & 8.20 & {\bf 7.23} & 7.44 & 7.46 \\
		Audikw\_1		& 70.9  & 64.2 & 66.5 & 60.1 & 58.7       & 55.8 & {\bf 55.0} & 59.2        & 56.3 & 55.9 \\
		Ieej			      & 6.98  & 6.49 & 6.72 & 5.34 & 6.93       & 6.42 & 5.18 & 5.97        & 5.51 & {\bf 4.83} \\
		\hline
		\end{tabular}

\vspace{0.3\baselineskip}

(c) Fujitsu CX2550 M4 (Intel Xeon Skylake)

\vspace{0.3\baselineskip}

\begin{tabular}{|c|c|ccc|ccc|ccc|}
		\hline
            \multirow{2}{*}{Dataset} & \multirow{2}{*}{MC} & \multicolumn{3}{|c}{BMC} & \multicolumn{3}{|c}{HBMC (crs\_spmv)} & \multicolumn{3}{|c|}{HBMC (sell\_spmv)} \\
			& & $b_{s}=8$ & $b_{s}=16$ & $b_{s}=32 $ & $b_{s}=8$ & $b_{s}=16$ & $b_{s}=32 $ & $b_{s}=8$ & $b_{s}=16$ & $b_{s}=32 $ \\ \hline
		Thermal2		      & 10.1  & 9.74        & 9.38 & 9.21 & {\bf 8.58} & 9.62 & 9.33 & 8.70        & 8.93 & 9.13 \\
		Parabolic\_fem	& 1.81  & 1.70       & 1.75 & 1.64 & {\bf 1.47}  & 1.57 & 1.78 & 1.52        & 1.49 & {\bf 1.49} \\
		G3\_circuit		& 6.17  & 5.07        & 5.23 & 5.01 & 4.83       & 5.19 & 5.60 & {\bf 4.54} & 5.09 & 4.87 \\
		Audikw\_1		& 44.6  & {\bf 34.6} & 36.4 & 36.6 & 35.5       & 37.5 & 37.9 & 37.4        & 37.3 & 39.8 \\
		Ieej			      & 3.76  & 4.71        & 3.72 & 3.56 & 4.20       & 3.69 & 3.37 & 3.81        & 3.51 & {\bf 3.15} \\
		\hline
		\end{tabular}

\end{table*}

\section{Related Works}
The parallelization of the sparse triangular solver for iterative solvers has been mainly investigated in the context of GS or IC/ILU smoothers for multigrid solvers, the SOR method, and IC/ILU preconditioned iterative solvers. Most of these parallelization techniques are classified into two classes: domain decomposition type methods and parallel orderings~\cite{duff2}.
A simple but commonly used technique in the former class is the additive Schwarz smoother or preconditioner.
The hybrid Jacobi and GS smoother, and block Jacobi IC/ILU preconditioning are typical examples, and they are used in many applications~\cite{local-ICCG,hybridsmoother}.
However, these techniques typically suffer from a decline in convergence when the number of threads (processes) is increased. Although there are some remedies for the deterioration in convergence, for example, the overlapping technique~\cite{radicati}, it is generally difficult to compensate for it when many cores are used.

A parallel ordering or reordering technique is a standard method to parallelize the sparse triangular solver.
We focus on the parallelization of IC/ILU preconditioners or smoothers; however, there are many studies that discuss the application of parallel ordering for GS and SOR methods, for example, \cite{intel2}. 
Ref.  \cite{duff2} provides an overview of early works on the parallel ordering method when applied to IC/ILU preconditioned iterative solvers.
Parallel orderings were mainly investigated in the context of a structured grid problem (a finite difference analysis), and the concepts of typical parallel ordering techniques, such as red-black, multi-color, zebra, domain decomposition (four or eight-corner), and nested dissection, were established in the 1990s.
In these early research activities, a major issue was the trade-off problem between convergence and the degree of parallelism.
After Duff and Meurant indicated the problem in \cite{Duff}, both analytical and numerical investigations were conducted in \cite{doi4, siam-iwa, doi3, Kuo, doi2, Eijkhout, doi1, benzi}.
The concept of equivalence of orderings and some remedies for the trade-off problem, such as the use of a relatively large number of colors in multi-color ordering or block coloring, were introduced as the results of these research activities.

In practical engineering and science domains, unstructured problems are solved more frequently than structured problems.
Therefore, parallel ordering techniques were enhanced for unstructured problems, and several heuristics were proposed.
Typical examples are hierarchical interface decomposition (HID)~\cite{henon} and heuristics for nodal or block multi-coloring~\cite{multi, amc, IPDPS2012}.
These techniques and other related methods have been used in various application domains, such as CFD, computational electromagnetics, and structure analyses \cite{semba, tsuburaya, CFD, kengo, nvidia}.    

Finally, we address the recently reported research results that are related to parallel linear solvers that involve sparse triangular solvers.
Gonzaga de Oliveira et al. reported their intensive numerical test results to evaluate various reordering techniques in the ICCG method in \cite{Oliveira}. 
Gupta introduced a blocking framework to generate a fast and robust preconditioner based on ILU factorization in \cite{gupta}. 
Chen et al. developed a couple of ILU-based preconditioners on GPUs in \cite{chen}. Ruiz et al. reported the evaluation results of HPCG implementations using nodal and block multi-color orderings on the ARM-based system, which confirmed the superiority of the block coloring method in \cite{arm}.

In this paper, we proposed a parallel ordering that is different from the techniques described above. To the best of our knowledge, there is no parallel ordering method that vectorizes the sparse triangular solver while maintaining the same convergence and number of synchronizations as the block multi-color ordering.
Since the vectorization of SpMV has been intensively investigated~\cite{sell-1,sell-2}, one of conventional approaches is the use of multi-color ordering, in which the substitution is represented as an SpMV in each color.
However, the multi-color ordering suffers from the problems of convergence and data locality, which are also indicated in the latest  report~\cite{arm}.
When we consider the numerical results and mathematical properties of the proposed hierarchical block multi-color ordering, it can be regarded as an effective technique for multithreading and vectorizing the sparse triangular solver.


\section{Conclusions}
In this paper, we proposed a new parallel ordering method, hierarchical block multi-color ordering (HBMC), for vectorizing and multithreading the sparse triangular solver. HBMC was designed to maintain the advantages of the block multi-color ordering (BMC) in terms of convergence and the number of synchronizations.
In the method, the coefficient matrix was transformed into the matrix with hierarchical block structures.
The level-1 blocks were mutually independent in each color, which was exploited in multithreading. 
Corresponding to the level-2 block,  the substitution was converted into $w$ (= SIMD width) independent steps, which were efficiently processed by SIMD instructions.
In this paper, we demonstrated analytically the equivalence of HBMC and BMC in convergence. 
Furthermore, numerical tests were conducted to examine the proposed method using five datasets on three types of computational nodes.
The numerical results also confirmed the equivalence of the convergence of HBMC and BMC.
Moreover, numerical tests indicated that HBMC outperformed BMC in 13 out 15 test cases (five datasets $\times$ three systems), which confirmed the effectiveness of the proposed method.
In the best case (G3\_circuit, Cray XC40), HBMC was 2.3 times faster than BMC.

In the future, we will examine our technique for other application problems, particularly a large-scale multigrid application and an HPCG benchmark.
Moreover, for further acceleration of the solver, we intend to introduce a sophisticated storage format or its related technique to our solver.
As other research issues, we will examine the effect of other coloring and blocking strategies on the performance of the solver.



\appendix
\section{Equivalent reordering}
In this paper, we only provide a sketch of the proof with respect to the ER condition because of a lack of space.
In the GS and SOR cases, the proof is given through the double use of the inductive method. The first is for the iteration step. In each iteration step, we use the inductive method again in a row-by-row manner. The proof is relatively straightforward.

In the IC/ILU preconditioning case, the proof consists of two parts. We first prove the equivalence of preconditioners $\bar{\LLL}=\PPP_{\pi} \LLL \PPP_{\pi}^{T}$ and $\bar{\UUU}=\PPP_{\pi} \UUU \PPP_{\pi}^{T}$.
We consider the right-looking factorization process with overwriting, and first prove that the updating process in the factorization is equivalent. Next, we show that the application of the updating process to $a_{i, j}$ is equivalent to that to $\bar{a}_{\pi(i), \pi(j)}$ (the $\pi(i)$-th row $\pi(j)$-th column element of $\bar{\AAA}$) when (\ref{equivalence}) holds. This leads to the equivalence of preconditioners. In the second part, the equivalence in the substitutions is shown using the inductive method, similar to the proof for the GS method.

\bibliographystyle{ACM-Reference-Format}

\end{document}